\documentclass[twoside]{article}
\usepackage[latin1]{inputenc}
\usepackage{t1enc}
\usepackage{a4}
\usepackage{tabularx}
\usepackage{epsf}

\usepackage{graphicx}

\def\bref{\vspace{4pt}\noindent\hangindent=10mm}

\newcommand{\Msun}{\ensuremath{M_{\odot}}}
\newcommand{\Zsun}{\ensuremath{Z_{\odot}}}

\newcommand{\Hb}{\ensuremath{{\rm H}\beta}}
\newcommand{\Mg}{\ensuremath{{\rm Mg}_2}}
\newcommand{\Mgb}{\ensuremath{{\rm Mg}\, b}}
\newcommand{\Fe}{\ensuremath{\langle {\rm Fe}\rangle}}
\newcommand{\Teff}{\ensuremath{T_{\rm eff}}}
\newcommand{\aFe}{\ensuremath{\alpha/{\rm Fe}}}
\newcommand{\FeH}{\ensuremath{{\rm Fe}/{\rm H}}}
\newcommand{\ZH}{\ensuremath{Z/{\rm H}}}

\begin{document}

\setcounter{figure}{0}
\setcounter{section}{0}
\setcounter{equation}{0}

\begin{center}
{\Large\bf
The Epochs of Early-Type Galaxy\\[0.2cm]
Formation in Clusters and in the Field}\\[0.7cm]

Daniel Thomas, Claudia Maraston, \& Ralf Bender \\[0.17cm]
Universit\"ats-Sternwarte M\"unchen \\
Scheinerstr.\ 1, D-81679 M\"unchen, Germany \\
daniel@usm.uni-muenchen.de
\end{center}

\vspace{0.5cm}

\begin{abstract}
\noindent{\it %
We compute new population synthesis models of Lick absorption line
indices with variable \aFe\ ratios and use them to derive average
ages, metallicities, and [\aFe] element enhancements for a sample of
126 field and cluster early-type galaxies. Calibrating the models on
galactic globular clusters, we show that any population synthesis
model being based on stellar libraries of the Milky Way is
intrinsically biased towards super-solar \aFe\ ratios at metallicities
below solar. We correct for this bias, so that the models presented
here reflect constant \aFe\ ratios at all metallicities. The use of
such unbiased models is essential for studies of stellar systems with
sub-solar metallicities like (extragalactic) globular clusters or
dwarf galaxies.

For the galaxy sample investigated here, we find a clear correlation
between [\aFe] and velocity dispersion. Zero-point, slope, and scatter
of this correlation turn out to be independent of the
environment. Additionally, the [\aFe] ratios and mean ages of
elliptical galaxies are well correlated, i.e.\ galaxies with high
\aFe\ ratios have also high average ages. This strongly reinforces the
view that the [\aFe] element enhancement in ellipticals is produced by
short star formation timescales rather than by a flattening of the
initial mass function. With a simple chemical evolution model, we
translate the derived average ages and \aFe\ ratios into star
formation histories. The more massive the galaxy, the shorter is its
star formation timescale, and the higher is the redshift of the bulk
of star formation, independent of the environmental density. We show
that this finding is incompatible with the predictions from
hierarchical galaxy formation models, in which star formation is
tightly linked to the assembly history of dark matter halos.  }
\end{abstract}

\section{Introduction}
The stellar population properties of early-type galaxies represent a
key challenge for theories of galaxy formation. In hierarchical models
of galaxy formation, more massive galaxies have longer assembly
timescales and therefore younger mean ages (Kauffmann 1996). Without
taking metallicity effects into account, this results in more massive
galaxies having bluer colors in contradiction to the observational
evidence (Bower, Lucey, \& Ellis 1992). It is well known, that
metallicity can be traded for age.  Therefore, by assuming that
metallicities steeply increase with the mass of the galaxy, the
blueing due to age effects can be masked by metallicity.  In this way
hierarchical galaxy formation models are able to produce the correct
slope of the color magnitude relation (Kauffmann \& Charlot 1998). It
remains open, however, if the correct ages of early-type galaxies are
predicted.  A meaningful test of the model must include the direct
comparison of predicted and observed ages.

A powerful tool to derive ages and metallicities are absorption line
indices (Worthey 1994). Under the assumption that the stellar
populations of ellipticals do not exhibit a significant spread in
metallicity (Maraston \& Thomas 2000), a combination of the Lick
indices \Hb, \Mgb, and $\Fe=({\rm Fe5270}+{\rm Fe5335})/2$ (Faber et
al.\ 1985) serves to disentangle age and metallicity. Still, this
approach is hampered by the fact that the data of elliptical galaxies
lie below the models in the \Mgb-\Fe-index diagram (e.g., Worthey,
Faber, \& Gonz\'alez 1992; Davies, Sadler, \& Peletier 1993). More
precisely, the Mg lines observed in elliptical galaxies are
stronger---at a given Fe line strength---than predicted by population
synthesis models. As a consequence, Mg line indices yield higher
metallicities and younger ages than the Fe line indices, which
indicates the presence of [\aFe] enhanced stellar populations. This
interpretation gets empirical support from the work of Maraston et
al. (2002), who find that metal-rich globular clusters of the Bulge
with independently known super-solar [\aFe] ratios exhibit the same
pattern in the \Mgb-\Fe\ and \Mg-\Fe\ diagrams.  A non-ambiguous
derivation of ages therefore requires population synthesis models with
variable \aFe\ ratios, and the consideration of the 3-dimensional
parameter space of Balmer, Mg, and Fe lines.

The paper is organized as follows. In Section~\ref{sec:models} we
present the main ingredients of our population synthesis models, whose
calibration on globular clusters is shown in
Section~\ref{sec:calibration}. The ages, metallicities, and \aFe\
ratios of early-type galaxies are derived and presented in
Section~\ref{sec:galaxies}. A comparison of these results with the
predictions from models of hierarchical galaxy formation is shown in
Section~\ref{sec:hc}. The main conclusions of this paper are discussed
and summarized in Sections~\ref{sec:discussion} and~\ref{sec:summary}.

\section{New Population Synthesis Models}
\label{sec:models}
The classical input parameters for population synthesis models are age
and metallicity. In this paper, we introduce the element abundance
ratio \aFe\ as a third input parameter. For the construction of the
$\alpha$/Fe-enhanced simple stellar population (SSP) models we use the
SSP models of Maraston (1998, 2002) as the base models, which we then
modify according to the desired $\alpha$/Fe ratio. In the following we
summarize the procedure and introduce the main input parameters. For a
more detailed presentation of the model we refer to Thomas, Maraston,
\& Bender (2002).

\subsection{The base SSP model}
The underlying solar-scaled SSP models are presented in Maraston
(1998, 2002). In these models, the fuel consumption theorem (Renzini
\& Buzzoni 1986) is adopted to evaluate the energetics of the post
main sequence phases. The input stellar tracks (solar abundance
ratios) with metallicities from 1/200 to 2 solar, are taken from
Cassisi, Castellani, \& Castellani (1997), Bono et al.\ (1997), and
S.~Cassisi (1999, private communication). The tracks with 3.5 solar
metallicity are taken from the solar-scaled set of Salasnich et al.\
(2000). Lick indices are computed by adopting the fitting functions of
Worthey et al.\ (1994).

\subsection{Element abundance variations}
We construct models with super-solar \aFe\ ratios by increasing the
abundances of the $\alpha$-elements (i.e.\ N, O, Mg, Ca, Na, Ne, S,
Si, Ti) and by decreasing those of the Fe-peak elements (i.e.\ Cr, Mn,
Fe, Co, Ni, Cu, Zn), such that total metallicity is conserved (Trager
et al.\ 2000a). The abundances of Carbon and all elements heavier than
Zinc are assumed not to vary. It is important to notice that
super-solar $\alpha$/Fe ratios at constant total metallicity are
accomplished mainly through a depletion of the Fe-peak element
abundances, because total metallicity is made up predominantly by
oxygen and the other $\alpha$-elements.

\subsection{Effects on absorption line indices}
\label{sec:response}
The abundance variations of individual elements in the stellar
atmospheres certainly impact on the observed absorption-line strengths
of a stellar population. This effect represents the principal
ingredient in the present $\alpha$/Fe-enhanced models. The variations
of the Lick absorption line indices owing to the element abundance
changes described in the previous section are taken from Tripicco \&
Bell (1995; hereafter TB95).

TB95 computed model atmospheres and synthetic spectra along a
5~Gyr-old isochrone with solar metallicity, alternately doubling the
abundances of the elements C, N, O, Mg, Fe, Ca, Na, Si, Cr, Ti, Mn,
Ni, and V. Note that total metallicity is not conserved but slightly
increased.  The impact of the abundance variations on the temperature
of the isochrone is on purpose not considered by TB95. All models are
based on the same isochrone with fixed $T_{\rm eff}$ and $\log g$
distributions, so that the abundance effects are isolated at a given
temperature and surface gravity.

On the model atmospheres defined by these parameters, TB95
measure the absolute Lick index value $I_0^{\rm TB95}$ and the index
change $\Delta I(i)^{\rm TB95}$, for the variation of the abundance of
element $i$. From these numbers one obtains the fractional index
change ({\em response function}) $R_{0.3}(X_i)=\Delta I(i)^{\rm
TB95}/I_0^{\rm TB95}$ of the index $I$ due to the enhancement of the
abundance of element $i$ by 0.3~dex.

Following Trager et al.\ (2000a), the total fractional change $\delta
I^{\rm TB95}$ of the index $I$ when enhancing all $\alpha$-elements
and depressing all Fe-peak elements can then be written as the product
of the fractional changes due to individual element abundance
variations:
\begin{equation}
\delta I^{\rm TB95} = \left\{ \prod_i [1+R_{0.3}(i)]^{([X_i/{\rm
H}]/0.3)}\right\} -1
\label{eqn:product}
\end{equation}
In this equation, $[X_i/{\rm H}]$ is the change of the abundance
ratios of element $i$ over Hydrogen relative to the solar value. This
equation assumes that the percentage index change is constant for each
step of 0.3 dex in abundance, which assures that index values approach
zero gracefully (Trager et al.\ 2000a).

The index variation $\Delta I$ of the index $I$ is then given by the
product of the total fractional index changes $\delta I^{\rm TB95}$
from TB95 and the value of the index $I$. Note, however, that $\delta
I^{\rm TB95}$ is calculated in TB95 for $Z=\Zsun$, so that this
procedure is in principle only valid for metallicities reasonably
close to solar. Indeed, with this prescription the $\alpha$/Fe ratios
of galactic globular clusters cannot be reproduced, the fractional
index changes given by TB95 turn out to be by far too small at
metallicities $Z\ll \Zsun$. As a first-order approximation, we
therefore assume that the {\em absolute} index change calculated by
TB95 at solar metallicity is conserved when going to lower
metallicities. At super-solar metallicities we adopt the {\em
fractional} index changes of TB95. The index variation $\Delta I$ for
the index $I$ for given element abundance variations $[X_i/{\rm H}]$
is then
\begin{equation}
\Delta I = \left\{
\begin{array}{lll}
\delta I^{\rm TB95}\times I & {\rm if}\ Z\geq\Zsun & ({\rm
fractional})\\[5pt] \delta I^{\rm TB95}\times I_0^{\rm TB95} & {\rm
if}\ Z<\Zsun & ({\rm absolute})
\end{array}
\right.
\label{eqn:responses}
\end{equation}

As response functions for metallicities different from solar are not
available, this is the most straightforward approximation at
present. It is further supported by the fact that the \aFe\ ratios we
derive for galactic globular clusters from their Lick indices \Mg,
\Fe, and \Hb\ are in very good agreement with independent
spectroscopic high-resolution measurements as shown in
Section~\ref{sec:calibration}.

It should also be emphasized that at metallicities relevant for
early-type galaxies and metal-rich globular clusters, i.e.\ $-0.5\leq
[\ZH]\leq 0.5$, the difference between fractional response and
absolute response in Eqn.~\ref{eqn:responses} has only a marginal
effect on the resulting SSP models (see Thomas et al.\ 2002).

\bigskip
TB95 compute the variations of the individual Lick indices on model
atmospheres with well defined values of temperature and gravity. These
values are chosen to be representative of the three evolutionary
phases dwarfs ($\Teff=4575~K$, $\log g=4.6$), turnoff ($\Teff=6200~K$,
$\log g=4.1$) and giants ($\Teff=4255~K$, $\log g=1.9$), on the 5~Gyr,
solar metallicity isochrone used by the authors.

We separate the turnoff region from the dwarfs on the Main Sequence at
5000~K, independent of age and metallicity. Note that the impact on
the final model from varying this temperature cut-off as a function of
age and metallicity is negligible (Thomas et al.\ 2002). We assign the
Sub Giant Branch phase to the turnoff because of the very similar
\Teff\ and $g$. The evolutionary phase `giants' consist of the Red
Giant Branch, the Horizontal Branch, and the Asymptotic Giant Branch
phases.  The Lick indices are computed for each evolutionary phase
separately, and modified according to the response functions presented
in Eqn.~\ref{eqn:responses}. The total integrated index of the SSP is
then
\begin{equation} 
I_{\rm SSP} = \frac{I^{\rm D} \times F_{\rm C}^{\rm D} + I^{\rm T}
\times F_{\rm C}^{\rm T} + I^{\rm G} \times F_{\rm C}^{\rm G}}{F_{\rm
C}^{\rm D} + F_{\rm C}^{\rm T} + F_{\rm C}^{\rm G}}\ ,
\label{eqn:sspus}
\end{equation}
where $I^{\rm D}$, $I^{\rm T}$, $I^{\rm G}$ are the integrated indices
of the three phases, and $F_{\rm C}^{\rm D}$, $F_{\rm C}^{\rm T}$,
$F_{\rm C}^{\rm G}$ are their continua. It can be easily verified that
Eqn.~\ref{eqn:sspus} is mathematically equivalent to
\begin{equation}
I_{\rm SSP} = \Delta \left(1-\frac{\sum_i F^i_{\rm L}}{\sum_i F^i_{\rm
C}}\right)\ ,
\label{eqn:ssp}
\end{equation}
which defines integrated indices (in EW) of SSPs. In Eqn.~\ref{eqn:ssp}
$F^i_{\rm L}$ and $F^i_{\rm C}$ are the fluxes in the line and the continuum
(of the considered index), for the $i$-th star of the population, $\Delta$
is the line width.

\subsection{\boldmath The $\alpha$/Fe bias of stellar libraries}
\label{sec:bias}
In population synthesis models, the link between Lick absorption line
indices and the stellar parameters temperature, gravity, and
metallicity is provided by stellar libraries that inevitably reflect
the chemical history of the Milky Way\footnote{In our (and most)
population synthesis models this link is given by the so-called
fitting functions of Worthey et al.\ (1994).}. This implies that every
population synthesis model suffers from a bias in the \aFe\ ratio,
i.e.\ the model does not reflect solar \aFe\ at all metallicities
(Borges et al.\ 1995). From the abundance patterns of the Milky Way
stars (see the review by McWilliam 1997 and references therein), we
know that the underlying \aFe\ ratio is solar at $Z=\Zsun$ and
increases with decreasing metallicity to $[\aFe]\approx 0.3$ at $Z\leq
\Zsun/10$.

\begin{table}[ht!]
\caption{The \aFe\ Bias in the Milky Way}
\begin{center}
\begin{tabular}{lrrrrrr}
\hline\hline
 {[\ZH]} & $-2.25$ & $-1.35$ & $-0.33$ & $0.00$ & $0.35$ & $0.67$\\
 {[\aFe]} & $0.40$ &  $0.30$ &  $0.10$ & $0.00$ & $0.00$ & $0.00$\\
\hline
\end{tabular}
\end{center}
\label{tab:bias}
\end{table}
In this paper we construct SSP models for different \aFe\ with the aid
of the index response functions as described in the previous
sections. With this procedure it is straightforward to correct for the
bias given by the chemical history of our Galaxy (see
Table~\ref{tab:bias}). The values in Table~\ref{tab:bias} are in
excellent agreement with the bias derived by Maraston et al.\ (2002)
with the aid of the fitting functions of Borges et al.\ (1995).  By
correcting for this bias we present for the first time SSP models that
have a constant \aFe\ ratio at all metallicities. The use of such
models is particularly important for the interpretation of metal-poor
stellar systems like (extragalactic) globular clusters or dwarf
spheroidal galaxies.

\subsection{\boldmath The effect of \aFe-enhanced stellar tracks}
In principle the element abundance variations in a star affect also
the star's evolution and the opacities in the stellar atmosphere,
hence the effective temperature. A fully consistent
$\alpha$/Fe-enhanced SSP model should therefore consider
$\alpha$-enhanced stellar evolutionary tracks. The impact of non-solar
$\alpha$/Fe ratios on stellar evolution, however, was very
controversially discussed in the literature (see discussion in Trager
et al.\ 2000a). In particular, we are still missing a homogeneous set
of stellar tracks with non-solar abundance ratios using consistent
opacities, that are computed for a large range in
metallicities. Salasnich et al.\ (2000) computed \aFe-enhanced stellar
tracks for metallicities $Z>1/2~\Zsun$, while Bergbusch \& VandenBerg
(2001) recently published \aFe-enhanced stellar tracks for
metallicities $Z<1/2~\Zsun$.  It would be valuable to have one set of
models that includes sub-solar metallicities for calibration purposes
on globular clusters and super-solar metallicities for the application
on early-type galaxies.

Nevertheless, in Thomas et al.\ (2002) we additionally compute models
in which the \aFe-enhanced stellar tracks of Salasnich et al.\ (2000)
are included at metallicities above 1/2 solar. The lower opacities of
the \aFe\ enhanced tracks lead to hotter isochrones, so that with
these models we derive unreasonably high ($\sim 30$~Gyr) ages for
early-type galaxies. Unfortunately, a calibration of the model with
the \aFe-enhanced tracks of Salasnich et al.\ (2000) on globular
clusters is not possible, because metallicities below 1/2 solar are
missing.  We therefore do not consider \aFe-enhanced stellar tracks in
the models presented here.

\section{Calibration on Globular Clusters}
\label{sec:calibration}
In this section we show the calibration of our \aFe-enhanced SSP
models on galactic globular clusters. We derive [\aFe] ratios and
metallicities from the \Mg\ and Fe5270/Fe5335 indices of the globular
clusters observed by Covino, Galletti, \& Pasinetti (1995). These are
then compared with independent spectroscopic measurements from
individual stars in these clusters, taken from the compilations by
Carney (1996) and Salaris \& Cassisi (1996).  We restrict our study to
the data of Covino et al.\ (1995), because they give very consistent
measurements of the indices Fe5270 and Fe5335 for all globulars
(except NGC~6356), which is not the case for the data of Trager et
al.\ (1998).

We do not use the \Hb\ index, but assume a fixed age of 12~Gyr for the
following reason. In Maraston, Greggio, \& Thomas (2001) it is shown
that, because of the appearance of hot horizontal branch stars at very
low metallicities, \Hb\ does not monotonically decrease with
increasing age for $[\FeH]\leq -1$, but has a minimum at $t\approx
12$~Gyr. As a consequence, the age determination through \Hb\ is
ambiguous at very low metallicities. Moreover, assuming the galactic
globular clusters to be uniformly old, their \Hb\ indices and in
particular the strong increase of \Hb\ with decreasing metallicity can
be perfectly reproduced with our models (Maraston \& Thomas 2000).
Note also that the galactic globular clusters are found to be coeval
independent of their metallicities (Rosenberg et al.\ 1999; Piotto et
al.\ 2000). The ages derived from color-magnitude diagrams lie between
9 and 14~Gyr (VandenBerg 2000).  The exact assumed age does not impact
on the \aFe\ ratios derived here, because of the mild derivatives with
age of the Mg and Fe indices at old ages.

\subsection{\boldmath \Mg\ and Fe indices}
\begin{figure}[ht!]
\centering\includegraphics[width=0.7\linewidth]{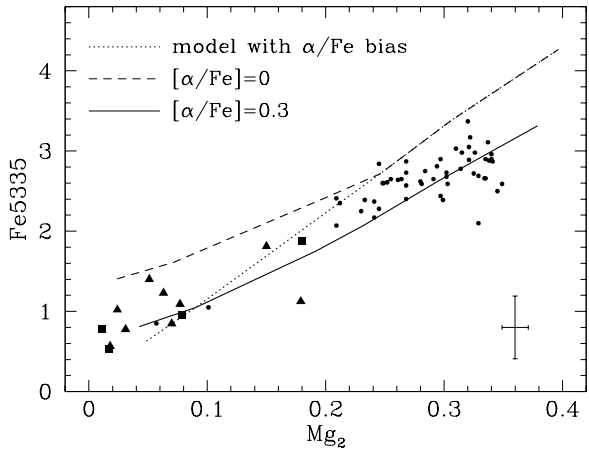}
\caption{ Index-index diagram Fe5335 vs.\ \Mg. Squares and triangles
are galactic globular cluster data from Covino et al.\ (1995).
Typical error bars for these data are shown in the lower-right corner
of the diagram.  Squares are those clusters for which independent
measurements of [\aFe] are given in Carney (1996) and Salaris \&
Cassisi (1996). The circles are early-type galaxies from Beuing et
al.\ (2002).  The lines are 12~Gyr SSP models in the metallicity range
$-2.25\leq [\ZH]\leq 0.67$. The dotted lines are the underlying SSP
models of Maraston (1998, 2002), which are biased in \aFe, i.e.\ they
reflect super-solar \aFe\ ratios at sub-solar metallicities (see
Section~\ref{sec:bias}).  The dashed and solid lines are the unbiased
SSP models of this paper for $[\aFe]=0.0$ and $[\aFe]=0.3$ at all
metallicities, respectively. }
\label{fig:mg2fe}
\end{figure}
Fig.~\ref{fig:mg2fe} shows data of galactic globular clusters from
Covino et al.\ (1995) as triangles and squares in the \Mg-Fe5335
plane. Squares are those clusters for which independent measurements
of [\aFe] are given in the literature.  SSP models of fixed age
(12~Gyr) and metallicities $-2.25\leq [\ZH]\leq 0.67$ are
over-plotted.  The dotted line are the underlying SSP models of
Maraston (1998, 2002). These models---like other SSP models in the
literature (Buzzoni, Gariboldi, \& Mantegazza 1992; Buzzoni,
Mantegazza, \& Gariboldi 1994; Worthey 1994; Tantalo et al.\ 1996;
Vazdekis et al.\ 1996; Kurth, Fritze-von Alvensleben, \& Fricke 1999;
and others)---are biased in \aFe, i.e.\ they reflect super-solar \aFe\
ratios at sub-solar metallicities (see
Section~\ref{sec:bias}). Therefore, at the lowest metallicities, the
models predict weaker Fe-indices than observed in globular
clusters. The dashed and solid lines are the unbiased SSP models of
this paper for $[\aFe]=0.0$ and $[\aFe]=0.3$ at all metallicities,
respectively. The globular cluster data lie between these two models,
indicating super-solar \aFe\ ratios in agreement with independent
spectroscopic measurements of single stars in these clusters (see next
section).

We note that it is very unlikely that the excess of Fe measured in the
globular clusters can be explained by anomalies in the horizontal
branch morphologies, i.e.\ clusters with larger Fe have relatively red
horizontal branches for their metallicities. The clusters considered
here do not show such an effect. The HBR parameter defined by Lee
(1990), which essentially is a measure for the fraction of
horizontal-branch stars on the blue side of the RR Lyrae region,
indicates blue horizontal branches for all metal-poor clusters in the
Covino et al.\ (1995) sample (Harris 1996).

\subsection{\boldmath The \aFe\ ratios}
\begin{figure}[ht!]
\begin{minipage}{0.49\linewidth}
\includegraphics[width=\linewidth]{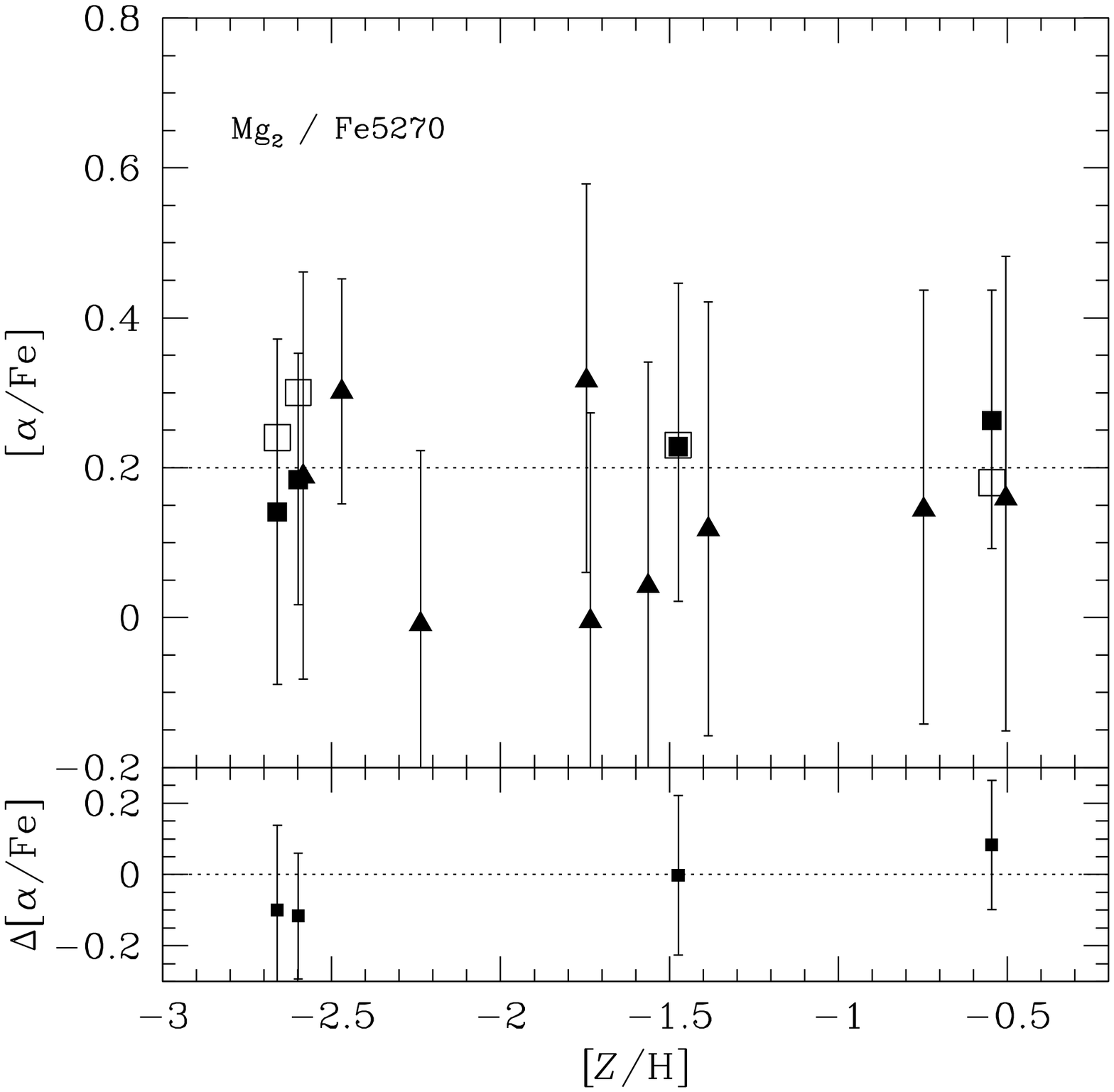}
\end{minipage}
\begin{minipage}{0.49\linewidth}
\includegraphics[width=\linewidth]{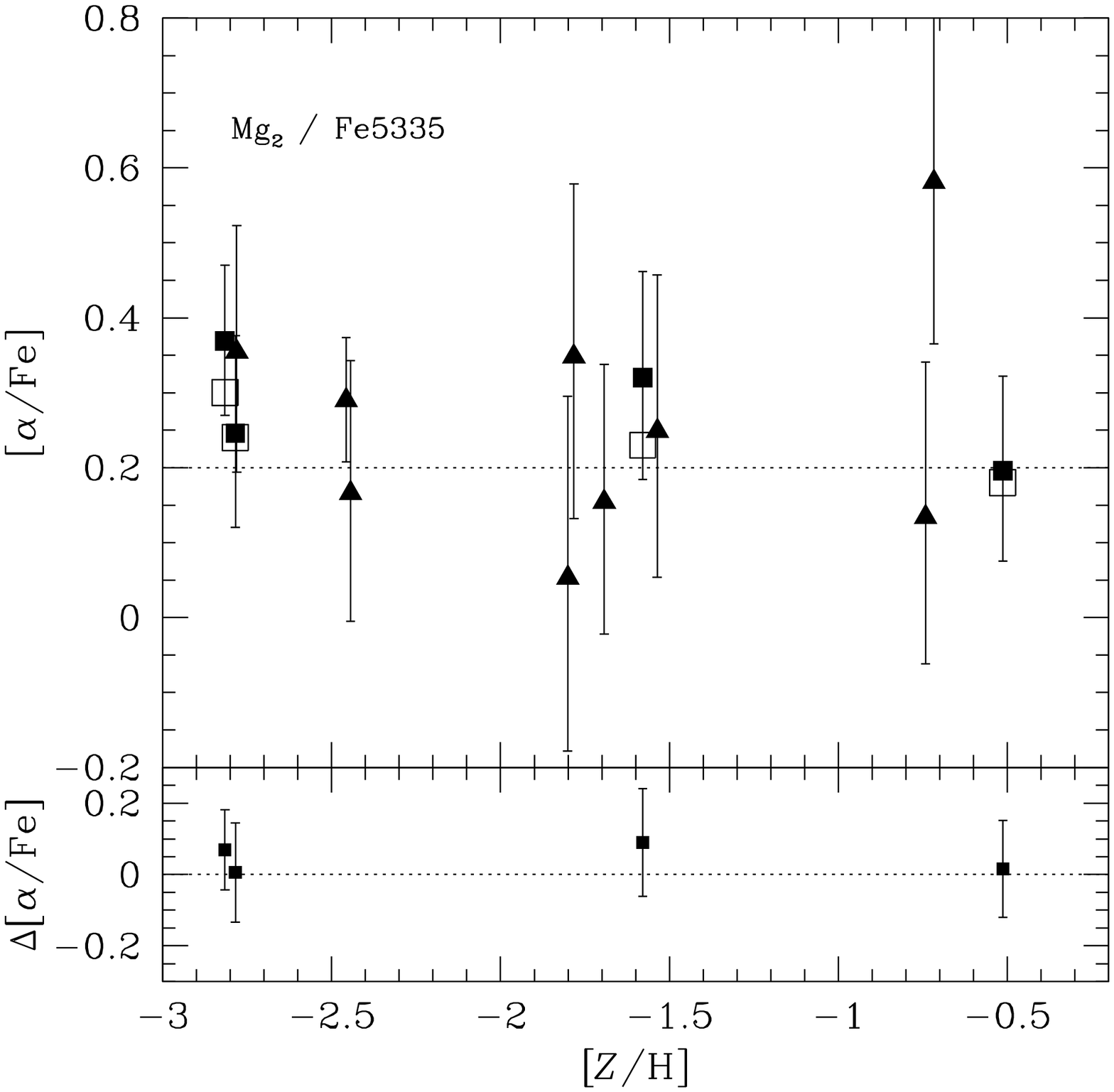}
\end{minipage}
\caption{Abundance ratios [\aFe] as a function of metallicity [\ZH]
for galactic globular clusters. Filled triangles and squares are the
Covino et al.\ (1995) sample for which we determined [\aFe] and [\ZH]
with our SSP models using \Mg\ and Fe5270 (left panel) or Fe5335
(right panel). Squares are those clusters for which independent
measurements of [\aFe] are given in Carney (1996) and Salaris \&
Cassisi (1996). The literature values are plotted as open squares.
The bottom panel shows the deviation of our determinations from the
literature: $\Delta [\aFe]=[\aFe]_{\rm This Work} - [\aFe]_{\rm
lit}$.}
\label{fig:zafe}
\end{figure}
In Fig.~\ref{fig:zafe} we plot the [\aFe] ratios derived for the
Covino et al.\ (1995) clusters versus their total metallicities [\ZH]
(filled symbols). Both these quantities are determined with our models
described in Section~\ref{sec:models}. The filled squares are those
clusters for which the [\aFe] ratios are known from independent
spectroscopic measurements of individual stars. These values are
plotted as open squares. The bottom panel shows the deviation of our
determinations from the literature: $\Delta [\aFe]=[\aFe]_{\rm This
Work} - [\aFe]_{\rm lit}$. The left and right panels show the results
for the Fe5270 and Fe5335 indices, respectively.

Within the errors, the values for [\aFe] derived with our models are
in good agreement with independent spectroscopic measurements in
single stars.  Both Fe indices yield consistent \aFe\ ratios, although
the errors (particular in Fe5270) are uncomfortably large.  A more
accurate calibration would certainly require data of better
quality. For the entire Covino et al.\ (1995) sample we find
$[\aFe]\approx 0.2-0.4$, independent of metallicity. This result
strongly supports the view that the galactic globular cluster
population formed early in a short ($<1$~Gyr) star formation episode,
so that no significant trend of age and \aFe\ with metallicity is
detectable.

For the application of the models to early-type galaxies a calibration
at solar and super-solar metallicities is desirable.  For a sample of
bulge clusters with metallicities up to solar (Maraston et al.\ 2002;
Puzia et al.\ 2002), we derive abundance ratios $0.2\leq [\aFe]\leq
0.4$, which are in good agreement with independent spectroscopic
element abundance measurements in these clusters (see Maraston et al.\
2002 and references therein).

\section{Key Parameters of Early-Type Galaxies}
\label{sec:galaxies}

\subsection{Data sample}
\label{sec:sample}
We analyze a sample of 126 early-type galaxies, 71 of which are field
and 55 cluster objects, containing roughly equal fractions of
elliptical and lenticular (S0) galaxies. The sample is constructed
from the following sources: 41 Virgo cluster and field galaxies
(Gonz\'alez 1993), 32 Coma cluster galaxies (Mehlert et al.\ 2000),
and 53 mostly field galaxies (highest quality data from Beuing et al.\
2002) selected from the ESO--LV catalog (Lauberts \& Valentijn
1989). In the latter sample, objects with a local galaxy surface
density NG$_{\mathrm{T}}>9$ are assumed to be cluster
galaxies. NG$_{\mathrm{T}}$ is given in Lauberts \& Valentijn (1989)
and is the number of galaxies per square degree inside a radius of one
degree around the considered galaxy.

\begin{figure}[ht!]
\begin{minipage}{\linewidth}
\includegraphics[width=\linewidth]{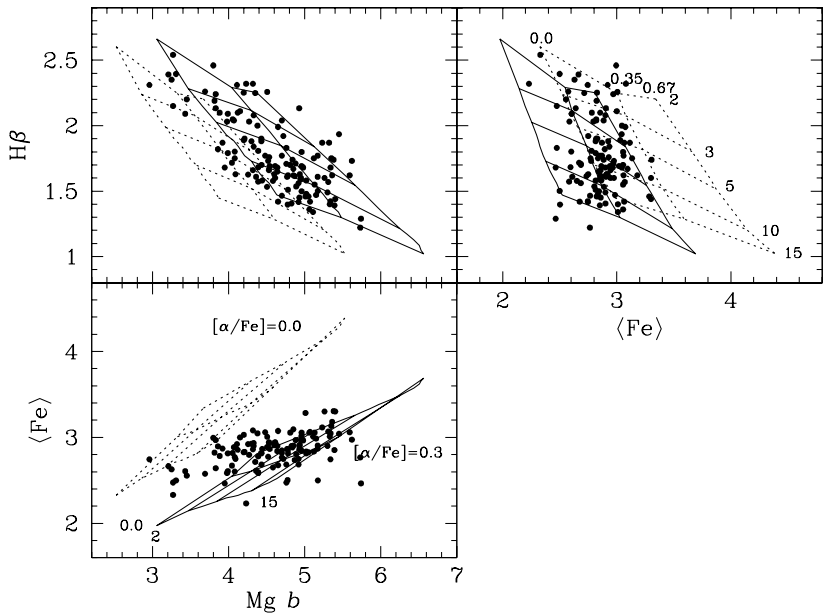}
\end{minipage}
\caption{Three-dim.\ parameter space of the Lick indices \Hb, \Mgb,
and $\Fe=({\rm Fe5270} + {\rm Fe5335})/2$. The filled circles are the
early-type galaxies analyzed in this paper, taken from Gonz\'alez
(1993), Mehlert et al.\ (2000), and Beuing et al.\ (2002).  Index
values are measured within $1/10\ r_e$.  Our models with $[\aFe]=0.0$
and $[\aFe]=0.3$ are plotted as dotted and solid lines,
respectively. Models of constant ages $t=2,3,5,10,15$~Gyr and
metallicities $[\ZH]=0.0,0.35,0.67$ are shown (see labels in the top
right, and bottom panels).}
\label{fig:grid}
\end{figure}
In Fig.~\ref{fig:grid} we show as filled circles the Lick absorption
line indices \Hb, \Mgb, and $\Fe=({\rm Fe5270}+{\rm Fe5335})/2$ of the
sample measured within $1/10\ r_e$. Overplotted are our models with
$[\aFe]=0.0$ as dotted lines and $[\aFe]=0.3$ as solid lines for the
metallicities $[\ZH]=0.0,0.35,0.67$ and for the ages
$t=2,3,5,10,15$~Gyr (see the labels in the top-right, and bottom
panels).

Note that the models with $[\aFe]=0.0$ are identical with the original
SSP models of Maraston (1998, 2002), as we do not assume any \aFe\
bias at these metallicities (see Table~\ref{tab:bias}). The \Mgb\
indices of the \aFe-enhanced model are higher, and the \Fe\ indices
are lower, while \Hb\ increases only marginally. At metallicities
close to solar, the fractional index changes are only very little
dependent on age and metallicity. Therefore, the shape of the model
grid is almost invariant under changes of \aFe.

By means of our \aFe-enhanced population synthesis models, we can now
uniquely determine the average ages, metallicities, and \aFe\ ratios
from the observed absorption indices \Hb, \Mgb, and \Fe.

\subsection{\boldmath The \aFe\ ratios}
\label{sec:relations}
\begin{figure}[ht!]
\includegraphics[width=\linewidth]{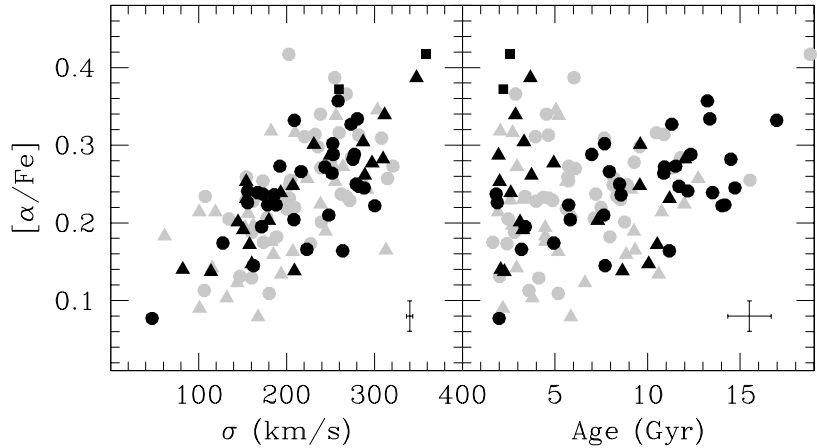}
\caption{Element abundance ratio [\aFe] as a function of velocity
dispersion $\sigma$ (measured within $1/10\ r_e$) and mean age. Grey
and black symbols are field and cluster early-type galaxies,
respectively. Triangles are lenticular, circles elliptical
galaxies. The two squares are the Coma cD galaxies NGC 4874 and NGC
4889. Mean ages and abundance ratios are derived with [\aFe] enhanced
SSP models described in Section~\ref{sec:models} from the indices \Hb,
\Mgb, and $\Fe=({\rm Fe5270}+{\rm Fe5335})/2$ measured within $1/10\
r_e$. Typical error-bars are given in the bottom-right corners. Galaxy
data are taken from Gonz\'alez (1993), Mehlert et al.\ (2000), and
Beuing et al.\ (2002).}
\label{fig:relations}
\end{figure}

In Fig.~\ref{fig:relations} we plot the element abundance ratio
[\aFe] as functions of velocity dispersion $\sigma$ (left panel) and
mean age (right panel). Grey and black symbols are field and cluster
early-type galaxies, respectively. Triangles are lenticular, circles
elliptical galaxies. Squares are the Coma cD galaxies NGC~4874 and
NGC~4889.

As anticipated qualitatively by Fisher, Franx, \& Illingworth (1995),
the [\aFe] ratio and $\sigma$ are well correlated, in agreement with
the study of Trager et al.\ (2000b). Additionally, we find that
zero-point, slope, and scatter of this relation are the same for field
and cluster galaxies.  This result extends the discovery that the
Mg-$\sigma$ relation is independent of environmental density (Bernardi
et al.\ 1998; Colless et al.\ 1999), and may provide a deeper
understanding of the origin for the Mg-$\sigma$ relation.

The increase of [\aFe] as a function of galaxy mass can be explained
by either a flattening of the initial mass function (IMF) or by a
shortening of the star formation timescale with increasing galaxy mass
(e.g., Matteucci 1994; Thomas, Greggio, \& Bender 1999). The
additional consideration of average ages helps to disentangle this
degeneracy. The right panel of Fig.~\ref{fig:relations} shows that the
[\aFe] element enhancement correlates with the average age of the
galaxy, such that objects with higher \aFe\ tend to be older. If IMF
variations were the main cause for the observed \aFe\ ratios, we would
not expect to find such a trend. In particular, the lack of old
objects with low \aFe\ in all environments could not be easily
understood.  This non-detection strongly supports the conclusion that
formation timescales rather than IMF variations are the driving
mechanism for the [\aFe]-$\sigma$ relation.

We conclude that the depth of the potential well, measured through the
velocity dispersion, defines the star formation timescales in
early-type galaxies and hence their \aFe\ ratios, independent of the
environment. If \aFe\ ratios are the main driver for the Mg-$\sigma$
relation, its independence of the environment is then easily
understood through the link between potential well and star formation
timescale.

From Fig.~\ref{fig:relations} it can be seen that the correlation
between mean [\aFe] and mean age is well defined for all cluster
ellipticals. For the two Coma cD galaxies and roughly 15 per cent of
the field ellipticals and lenticular galaxies we derive ages below
5~Gyr and $[\aFe]\geq 0.3$. Naively interpreted, the coexistence of
low ages and high \aFe\ could imply that the majority of stars in
these objects were formed recently on a short timescale. However, this
is very implausible. More likely explanations are:
\begin{itemize}
\item Metal-poor subcomponent. Old, metal-poor stellar populations
develop hot horizontal branches, which lead to rather strong Balmer
absorption despite the old age. A composite stellar population that
includes a small fraction of metal-poor stars can explain the Balmer
absorption of these objects ($\Hb\approx 2$) without invoking young
ages (Maraston \& Thomas 2000). In this case the derivation of young
average ages is an artifact, the galaxy is about 15~Gyr old, formed
its stars on a short timescale and is therefore \aFe-enhanced.
\item Recent star formation. The object formed most of its stars at
high redshift and suffered very recently ($\sim 300$~Myr ago) from a
minor (a few per cent in mass) star formation episode. As a
consequence, the object has a very young $V$-light averaged age, but
also a very low average \aFe\ because of the Fe enrichment from Type
Ia supernovae of the underlying old population (Thomas et al.\
1999). In this case, the seemingly large \aFe\ derived here is an
artifact, as composite stellar populations with a major old and a very
small very young subcomponent can mimic the existence of
\aFe-enhancement (Kuntschner 2000).  This effect comes from the
different partial time derivatives of the Mg and the Fe indices at
ages below 1~Gyr. We note the caveat, however, that the fitting
functions, and therefore also the population synthesis models, are not
valid at such low ages (Buzzoni et al.\ 1994; Worthey et al.\ 1994).
\end{itemize}

\subsection{The ages}
The correlation of \aFe\ with both $\sigma$ and average
age--essentially valid for elliptical galaxies---implies a relation
between average age and velocity dispersion. Both in clusters and in
the field, more massive objects are older.
\begin{figure}[ht!]
\centering\includegraphics[width=0.7\linewidth]{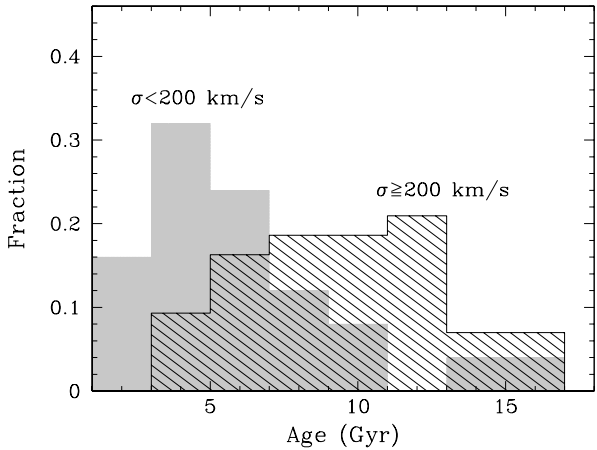}
\caption{Age distributions of elliptical galaxies with velocity
dispersions above (shaded histogram, 43 objects) and below (grey
histogram, 25 objects) 200 km/s. The histograms include both cluster
and field galaxies.}
\label{fig:ages}
\end{figure}

This result is better illustrated in Fig.~\ref{fig:ages}, in which the
age distributions of elliptical galaxies with velocity dispersions
above and below 200 km/s are shown as the shaded and grey histograms,
respectively.  More than 50 per cent of the low-mass ellipticals, but
only 10 per cent of the massive ellipticals, have average ages younger
than 5~Gyr.  The trend of massive ellipticals being older is in
agreement with a recent study by Poggianti et al.\ (2001a) of a large
number of Coma cluster galaxies. We add the important conclusion that
this correlation is an intrinsic property of elliptical galaxies and
does not depend on the environment.

It is interesting to note that lenticular galaxies do not follow this
trend. There is a considerable fraction of S0 galaxies with $\sigma>
200$~km/s, for which we derive relatively young average ages
($<5$~Gyr). This might indicate recent star formation episodes in such
objects as concluded by Poggianti et al.\ (2001b) or the existence of
metal-poor subpopulations (Maraston \& Thomas 2000; see the previous
section).

\subsection{Star formation histories}
The relations shown in Fig.~\ref{fig:relations} can be used to
constrain the epochs of the main star formation episode and the star
formation timescales for objects as a function of their velocity
dispersions. Assuming a Gaussian distribution for the star formation
rate, we calculate the chemical enrichment of $\alpha$ and Fe-peak
elements for an initially primordial gas cloud.  The delayed
enrichment from Type Ia supernovae is taken into account using the
prescription of Greggio \& Renzini (1983; see Thomas, Greggio, \&
Bender 1998, 1999 for more details).  The simulations are done for a
set of different star formation histories (Gaussians) varying the star
formation timescale (width) and the lookback time of the maximum star
formation (peak). For this set of star formation histories we compute
the $V$-light averaged ages and [\aFe] ratios of the resulting
composite stellar population today. These can be directly compared
with the observationally derived values plotted in
Fig.~\ref{fig:relations}.  Finally, the Gaussians are linked to
velocity dispersion using the relation between $\sigma$ and [\aFe]
(left panel of Fig.~\ref{fig:relations}).

\begin{figure}[t]
\centerline{\includegraphics[width=0.7\linewidth]{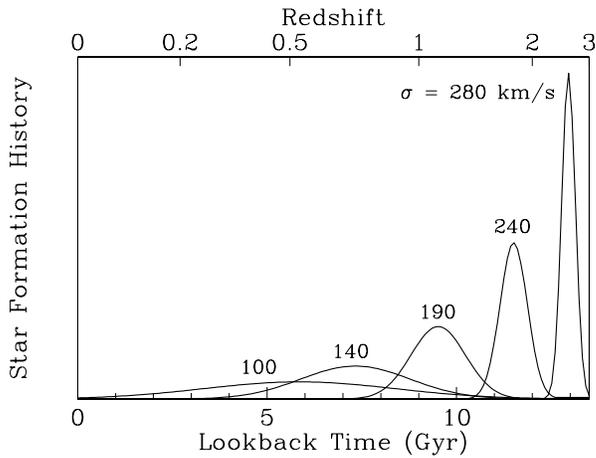}}
\caption{Star formation rates as functions of lookback time for
early-type galaxies with different velocity dispersions $\sigma$
between 100 km/s and 300 km/s. The star formation histories are
derived from the mean ages and [\aFe] ratios shown in
Fig.~\ref{fig:relations} (see text).  Redshifts assume $\Omega_m=0.2,\
\Omega_{\Lambda}=0.8,\ H_0=65$ km/s/Mpc.}
\label{fig:sfr}
\end{figure}
The resulting star formation rates as a function of lookback time are
shown in Fig.~\ref{fig:sfr}. The upper y-axis gives the corresponding
redshifts assuming $\Omega_m=0.2,\ \Omega_{\Lambda}=0.8$, and
$H_0=65$~km/s/Mpc. The more massive the galaxy, the shorter is its
formation timescale and the higher is its formation redshift.
Galaxies with $\sigma<140$~km/s exhibit significant star formation at
redshifts $z<1$. The star formation histories shown in
Fig.~\ref{fig:sfr} are in principle derived for cluster ellipticals,
but are also valid for the bulk of field ellipticals and S0 galaxies,
as only roughly 15 per cent deviate from the [\aFe]-age relation in
Fig.~\ref{fig:relations}.

Most interestingly, objects are observed at high redshifts whose
properties are consistent with the abundance ratios and star formation
histories derived in this paper. At least part of the SCUBA sources at
$2\leq z\leq 3$ turn out to be star forming galaxies with extremely
high star formation rates up to 1000~\Msun/yr (de Mello et al.\ 2002;
Smail et al.\ 2002). They are difficult to detect in the optical
rest-frame because they are enshrouded in dust. These objects are
likely to be the precursors of the most massive ellipticals
($\sigma\approx 300$~km/s) forming in a violent star formation episode
at high redshift. Ly-break galaxies, instead, exhibit more moderate
star formation rates of the order a few 10~\Msun/yr (Pettini et al.\
2001), and may therefore be the precursors of less massive ellipticals
($\sigma\approx 200$~km/s). In a recent work, Pettini et al.\ (2002)
analyze deep Keck spectra of the lensed Ly-break galaxy cB58 at
redshift $z=2.73$ (Seitz et al.\ 1998). They find the {\em
interstellar medium} to be significantly \aFe-enhanced, which is in
good agreement with the \aFe-enhancement derived in this paper for the
{\em stellar population} of local elliptical galaxies.

\section{Hierarchical Galaxy Formation}
\label{sec:hc}

The element abundance ratios $0.2\leq [\aFe]\leq 0.4$ derived in this
paper for massive ($\sigma>200$~km/s) early-type galaxies require star
formation timescales below $\sim 1$~Gyr (see Fig.~\ref{fig:sfr}). In
models of hierarchical galaxy formation, however, star formation in
ellipticals typically does not truncate after 1~Gyr, but continues to
lower redshift (Kauffmann 1996). It is therefore questionable if
hierarchical clustering would lead to significantly \aFe-enhanced
giant ellipticals (Bender 1996). In a more quantitative investigation,
Thomas et al.\ (1999) compute the chemical enrichment in mergers of
evolved spiral galaxies, and show that such a merger does not produce
significantly super-solar \aFe\ ratios. This prediction is confirmed
observationally by the study of Maraston et al.\ (2001), who find that
the newly formed globular clusters in the merger remnant NGC~7252 and
the integrated light of the galaxy are indeed not \aFe-enhanced.

So far, semi-analytic models have not considered this
constraint. Thomas (1999) demonstrates that the average star formation
history of cluster ellipticals within the hierarchical formation
scheme leads to $[\aFe]\sim 0.04$, a value that clearly lies below the
observational estimates in Fig.~\ref{fig:relations}. In Thomas \&
Kauffmann (1999), we follow this aspect in more detail by taking the
star formation histories of individual Monte Carlo realizations of
elliptical galaxies into account. This allows us to explore the
distribution and the scatter of the \aFe\ ratios among the galaxies as
they are predicted by the semi-analytic models. We calculate the
chemical enrichment of closed box systems for the star formation rate
predicted by semi-analytic models. Both modes of star formation, the
quiescent and the merger induced burst components, are taken into
account.  A universal Salpeter IMF slope $x=1.35$ is assumed.  The
simulations in this analysis are for a cold dark matter power spectrum
with $\Omega=1$, $H_0=50$~km/s/Mpc, and $\sigma_8=0.67$.  For more
details please refer to Thomas \& Kauffmann (1999) and Thomas (1999).

\begin{figure}[ht!]
\centering\includegraphics[width=0.7\linewidth]{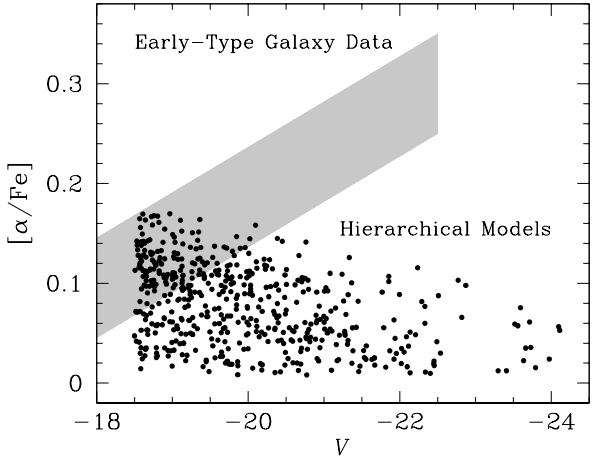}
\caption{Absolute $V$-magnitude vs.\ [\aFe] ratio. Points are
early-type galaxies modeled in the framework of hierarchical galaxy
formation (Thomas \& Kauffmann 1999). Halos with circular velocities
$V_c=1000$~km/s are considered, corresponding to cluster environments.
The grey shaded area indicates the range of [\aFe] values derived in
this paper from observed absorption line indices (see
Fig.~\ref{fig:relations}).}
\label{fig:hc}
\end{figure}
In Fig.~\ref{fig:hc}, we plot the resulting [\aFe] ratios of
elliptical galaxies as a function of their $V$-magnitudes. The
observationally derived [\aFe] ratios from Fig.~\ref{fig:relations}
are indicated as the grey shaded area.  The scatter in \aFe\ predicted
by the hierarchical models is large because of the relatively large
variety of formation histories comprising star formation timescales
from $10^9$ to $10^{10}$ yr.

The figure reveals a striking failure of the hierarchical model: More
luminous ellipticals are predicted to have lower \aFe\ ratios. The
opposite is observed. As a consequence, the \aFe\ ratios predicted by
the hierarchical models for massive elliptical galaxies are
significantly below the observed values. These problems are directly
related to the hierarchical clustering scheme, in which the largest
objects form last and have therefore more extended star formation
histories. This hierarchical paradigm leads also to the prediction of
an age-mass anti-correlation (Kauffmann \& Charlot 1998), which again
stands in clear contradiction to the observational evidence that
early-type galaxies with larger velocity dispersions tend to have
higher average ages (see Section~\ref{sec:relations}).

\section{Discussion}
\label{sec:discussion}
As recently summarized by Peebles (2002), the arguments of this paper
are accompanied by a number of evidences from the tightness and
redshift evolution of the scaling relations of early-type
galaxies. Both push the formation ages of the stellar populations in
large elliptical galaxies to redshifts $z>2$ (e.g., Bower et al.\
1992; Bender, Burstein, \& Faber 1993; Renzini \& Ciotti 1993; Bender,
Ziegler, \& Bruzual 1996; Kodama, Bower, \& Bell 1999; Ziegler et al.\
1999). These findings are not accomplished by standard hierarchical
models.

At least part of the reason for this failure may be connected to the
way how non-baryonic dark matter and baryonic matter are linked. In
current models, star formation is tightly linked to the assembly
history of dark matter halos, so that galaxies with longer assembly
times also form stars on longer timescales. In a recent paper, Granato
et al.\ (2001) present a promising---even though more
heuristic---approach, in which star formation is enhanced in massive
systems. The resulting 'anti-hierarchical baryonic collapse' leads to
higher formation redshifts and shorter formation timescales of the
stellar populations in massive objects. A modification of this kind is
certainly the right step towards reconciling hierarchical models with
the constraints set by the stellar populations properties of
early-type galaxies.  Still, this modification may not be sufficient
to harmonize hierarchical galaxy formation with the conclusion of
Gerhard et al.\ (2001), who argue that the very high core densities of
the halos of elliptical galaxies push also the collapse of their dark
matter halos (not only of their baryonic matter) to high redshifts.

\section{Summary}
\label{sec:summary}
In this paper we derive the central average ages, metallicities, and
\aFe\ ratios for 126 early-type galaxies from both cluster and field
environments.

For this purpose we develop new population synthesis models with
variable \aFe\ ratios, that allow for the derivation of these stellar
population parameters from the Lick absorption line indices \Hb, \Mgb,
and \Fe. The models are based on the SSP models of Maraston (1998,
2002). The effect from varying the \aFe\ ratio is calculated with the
Tripicco \& Bell (1995) index response functions, following the method
introduced by Trager et al.\ (2000a). At sub-solar metallicities we
calibrate our models on galactic globular clusters, putting particular
attention to the derivation of \aFe\ ratios. The base SSP models of
Maraston (1998, 2002)---like all population synthesis models up to
now---rely on stellar libraries that reflect the chemical history of
the Milky Way. They are therefore biased towards super-solar \aFe\
ratios at sub-solar metallicities. We correct for this bias, so that
our new SSP models reflect constant \aFe\ ratios at all metallicities.

For the sample of early-type galaxies investigated here, we find that
\aFe\ ratios correlate well with velocity dispersion $\sigma$,
independent of the environmental density. Elliptical galaxies and the
majority of lenticular galaxies additionally exhibit a good
correlation of \aFe\ ratio with average age. These results strongly
support the view that the increase of \aFe\ with increasing velocity
dispersion is due to a decrease of star formation timescales rather
than due to a flattening of the IMF. We show that more massive
ellipticals have higher average ages and higher \aFe\ ratios, because
of earlier formation epochs and shorter formation timescales of their
stellar populations. This observational result is not matched by
current models of hierarchical galaxy formation, mainly because they
predict too extended star formation histories for massive ellipticals.

\subsection*{Acknowledgments}
DT thanks the organizers of the JENAM 2001 conference for the
invitation to a highlight talk. We would like to acknowledge Santi
Cassisi for providing a large number of stellar evolutionary tracks,
and Beatriz Barbuy, Laura Greggio, Claudia Mendes de Oliveira, D\"orte
Mehlert, and Manuela Zoccali for very interesting and stimulating
discussions. DT and CM give sincere thanks to Claudia Mendes de
Oliveira, Beatriz Barbuy, and the members of the Instituto Astronomico
e Geofisico of S\~ao Paulo for their kind hospitality.  The the BMBF,
the DAAD, and the "Sonderforschungsbereich 375-95 f\"ur
Astro-Teilchenphysik" of the DFG are acknowledged for financial
support.

\subsection*{References}

{\small

%\bref
%Smith, H.\,D., Miller, P. 1999, A\&A 355, 123 %for example

\bref
Bender, R. 1996, in New Light on Galaxy Evolution, ed. R.~Bender \& R.\,L.
  Davies (Dordrecht: Kluwer Academic Publishers), 181

\bref
Bender, R., Burstein, D.,  \& Faber, S.\,M. 1993, ApJ, 411, 153

\bref
Bender, R., Ziegler, B.\,L.,  \& Bruzual, G. 1996, ApJ, 463, L51

\bref
Bergbusch, P.\,A.,  \& VandenBerg, D.\,A. 2001, ApJ, 556, 322

\bref
Bernardi, M.,  et~al. 1998, ApJ, 508, 143

\bref
Beuing, J., Bender, R., Mendes~de Oliveira, C., Thomas, D.,  \& Maraston, C.
  2002, A\&A, submitted

\bref
Bono, G., Caputo, F., Cassisi, S., Castellani, V.,  \& Marconi, M. 1997, ApJ,
  489, 822

\bref 
Borges, A.\,C., Idiart, T.\,P., de~Freitas~Pacheco, J.\,A., \&
Th{\'e}venin, F.  1995, AJ, 110, 2408

\bref
Bower, R.\,G., Lucey, J.\,R.,  \& Ellis, R.\,S. 1992, MNRAS, 254, 589

\bref
Buzzoni, A., Gariboldi, G.,  \& Mantegazza, L. 1992, AJ, 103, 1814

\bref
Buzzoni, A., Mantegazza, L.,  \& Gariboldi, G. 1994, AJ, 107, 513

\bref
Carney, B.\,W. 1996, PASP, 108, 900

\bref
Cassisi, S., Castellani, M.,  \& Castellani, V. 1997, A\&A, 317, 10

\bref
Colless, M., Burstein, D., Davies, R.\,L., McMahan, R.\,K., Saglia, R.\,P.,  \&
  Wegner, G. 1999, MNRAS, 303, 813

\bref
Covino, S., Galletti, S.,  \& Pasinetti, L.\,E. 1995, A\&A, 303, 79

\bref
Davies, R.\,L., Sadler, E.\,M.,  \& Peletier, R.\,F. 1993, MNRAS, 262, 650

\bref
de~Mello, D., Wiklind, T., Leitherer, C.,  \& Pontoppidan, K. 2002, in The
  Evolution of Galaxies II. Basic Building Blocks, ed. M.~Sauvage,
  G.~Stasinska, L.~Vigroux, D.~Schaerer, \& S.~Madden (Dordrecht: Kluwer), in
  press

\bref
Faber, S.\,M., Friel, E.\,D., Burstein, D.,  \& Gaskell, D.\,M. 1985, ApJS, 57,
  711

\bref
Fisher, D., Franx, M.,  \& Illingworth, G. 1995, ApJ, 448, 119

\bref
Gerhard, O., Kronawitter, A., Saglia, R.\,P., \& Bender, R. 2001, AJ,
121, 1936

\bref
Gonz{\'{a}}lez, J. 1993, Phd~thesis, University of California, Santa Cruz

\bref
Granato, G.\,L., Silva, L., Monaco, P., Panuzzo, P., Salucci, P.,
  De~Zotti, G., \& Danese, L. 2001, MNRAS, 324, 757

\bref
Greggio, L.,  \& Renzini, A. 1983, A\&A, 118, 217

\bref
Harris, W.\,E. 1996, AJ, 112, 1487

\bref
Kauffmann, G. 1996, MNRAS, 281, 487

\bref
Kauffmann, G.,  \& Charlot, S. 1998, MNRAS, 294, 705

\bref
Kodama, T., Bower, R.\,G.,  \& Bell, E.\,F. 1999, MNRAS, 306, 561

\bref
Kuntschner, H. 2000, MNRAS, 315, 184

\bref
Kurth, O.\,M., Fritze-v. Alvensleben, U., \& Fricke, K.\,J. 1999,
A\&AS, 138, 19

\bref
Lauberts, A.,  \& Valentijn, E.\,A. 1989, The Surface Photometry Catalogue of
  the ESO--Upsalla Galaxies (Garching: ESO)

\bref
Lee, Y.-W. 1990, ApJ, 363, 159

\bref
Maraston, C. 1998, MNRAS, 300, 872

\bref
Maraston, C. 2002, MNRAS, in preparation

\bref
Maraston, C.,  et~al. 2002, in preparation

\bref
Maraston, C., Greggio, L.,  \& Thomas, D. 2001, Ap\&SS, 276, 893

\bref
Maraston, C., Kissler-Patig, M., Brodie, J.\,P., Barmby, P.,  \& Huchra, J.
  2001, A\&A, 370, 176

\bref
Maraston, C.,  \& Thomas, D. 2000, ApJ, 541, 126

\bref
Matteucci, F. 1994, A\&A, 288, 57

\bref
McWilliam, A. 1997, ARA\&A, 35, 503

\bref
Mehlert, D., Saglia, R.\,P., Bender, R.,  \& Wegner, G. 2000, A\&AS, 141, 449

\bref
Peebles, P.\,J.\,E. 2002, in A New Era in Cosmology, ed. N.~Metcalfe \&
  T.~Shanks, ASP Conference Series (Dordrecht: Kluwer Academic Publishers), in
  press

\bref
Pettini, M., Rix, S.\,A., Steidel, C.\,C., Adelberger, K.\,L., Hunt,
  M.\,P., \& Shapley, A.\,E. 2002, ApJ, in press, astro-ph/0110637

\bref
Pettini, M., Shapley, A.\,E., Steidel, C.\,C., Cuby, J., Dickinson, M.,
  Moorwood, A.\,F.\,M., Adelberger, K.\,L.,  \& Giavalisco, M. 2001, ApJ,
  554, 981

\bref
Piotto, G., Rosenberg, A., Saviane, I., Zoccali, M., \& Aparicio,
  A. 2000, in Ap{\&}SS Library, Vol. 255, The evolution of the Milky
  Way: stars versus clusters, ed. F.~Matteucci \& F.~Giovannelli
  (Dordrecht: Kluwer Academic Publishers), 249

\bref
Poggianti, B.,  et~al. 2001a, ApJ, 562, 689

\bref
Poggianti, B., et~al. 2001b, ApJ, 563, 118

\bref
Puzia, T.,  et~al. 2002, in preparation

\bref
Renzini, A.,  \& Buzzoni, A. 1986, in Spectral evolution of galaxies, ed.
  C.~Chiosi \& A.~Renzini (Dordrecht: Reidel), 135

\bref
Renzini, A.,  \& Ciotti, L. 1993, ApJ, 416, L49

\bref
Rosenberg, A., Saviane, I., Piotto, G.,  \& Aparicio, A. 1999, AJ, 118, 2306

\bref
Salaris, M.,  \& Cassisi, S. 1996, A\&A, 305, 858

\bref
Salasnich, B., Girardi, L., Weiss, A.,  \& Chiosi, C. 2000, A\&A, 361, 1023

\bref
Seitz, S., Saglia, R.\,P., Bender, R., Hopp, U., Belloni, P.,  \& Ziegler, B.
  1998, MNRAS, 298, 945

\bref
Smail, I., Ivison, R.\,J., Blain, A.\,W., \& Kneib, J.-P. 2002, MNRAS,
  in press, astro-ph/0112100

\bref
Tantalo, R., Chiosi, C., Bressan, A.,  \& Fagotto, F. 1996, A\&A, 311, 361

\bref
Thomas, D. 1999, MNRAS, 306, 655

\bref
Thomas, D., Greggio, L.,  \& Bender, R. 1998, MNRAS, 296, 119

\bref
Thomas, D., Greggio, L.,  \& Bender, R. 1999, MNRAS, 302, 537

\bref
Thomas, D.,  \& Kauffmann, G. 1999, in Spectrophotometric dating of stars and
  galaxies, ed. I.~Hubeny, S.~Heap, \& R.~Cornett, Vol. 192 (ASP Conf. Ser.),
  261

\bref
Thomas, D., Maraston, C.,  \& Bender, R. 2002, MNRAS, in preparation

\bref
Trager, S.\,C., Faber, S.\,M., Worthey, G., \& Gonz{\'{a}}lez,
  J.\,J. 2000a, AJ, 119, 164

\bref
Trager, S.\,C., Faber, S.\,M., Worthey, G., \& Gonz{\'{a}}lez,
  J.\,J. 2000b, AJ, 120, 165

\bref
Trager, S.\,C., Worthey, G., Faber, S.\,M., Burstein, D., \&
  Gonz{\'{a}}lez, J.\,J. 1998, ApJS, 116, 1

\bref
Tripicco, M.\,J.,  \& Bell, R.\,A. 1995, AJ, 110, 3035

\bref
VandenBerg, D.\,A. 2000, ApJS, 129, 315

\bref
Vazdekis, A., Casuso, E., Peletier, R.\,F., \& Beckmann, J.\,E. 1996,
  ApJS, 106, 307

\bref
Worthey, G. 1994, ApJS, 95, 107

\bref
Worthey, G., Faber, S.\,M.,  \& Gonz{\'{a}}lez, J.\,J. 1992, ApJ, 398, 69

\bref
Worthey, G., Faber, S.\,M., Gonz{\'{a}}lez, J.\,J., \& Burstein,
  D. 1994, ApJS, 94, 687

\bref

Ziegler, B.\,L., Saglia, R.\,P., Bender, R., Belloni, P., Greggio, L.,
  \& Seitz, S. 1999, A\&A, 346, 13

}

\vfill

\end{document}